\documentclass[a4paper]{amsart}
\def\mylabelonoff{off}
\def\allowdisbrkyesno{yes}
\def\numberingtheoremsectionyesno{no}
\def\numberingequationsectionyesno{no}
\def\pagesizeextendednormal{extended}
\usepackage[mathscr]{eucal}
\usepackage[english]{babel}
\usepackage{a4,exscale,ifthen,amsfonts,amssymb,amsmath,amscd,graphicx,color}
\usepackage{nicefrac,tikz,fancyhdr,caption,array,multirow,multicol,booktabs,algorithm,algorithmic}
\usepackage[all]{xy}

\ifthenelse{\equal{\mylabelonoff}{on}}
{\newcommand{\mylabel}[1]{\label{#1}\fbox{{\sf #1}}}}
{\newcommand{\mylabel}[1]{\label{#1}}}
\ifthenelse{\equal{\allowdisbrkyesno}{yes}}
{\allowdisplaybreaks}
{}

\ifthenelse{\equal{\pagesizeextendednormal}{extended}}
{\setlength{\textwidth}{16cm}
\setlength{\textheight}{22cm}
\setlength{\oddsidemargin}{-0.2cm}
\setlength{\evensidemargin}{-0.2cm}}
{}
\ifthenelse{\equal{\numberingequationsectionyesno}{yes}}
{\numberwithin{equation}{section}}
{}

\newcommand{\cred}{\color{red}}

\newcommand{\diss}{\displaystyle}

\newcommand{\extended}[1]{{\cred #1}}
\newcommand{\condensed}[1]{}

\newcommand{\ccz}{c_{\mathsf{CZ}}}

\newcommand{\set}[2]{\{#1\,:\,#2\}}
\newcommand{\setb}[2]{\big\{#1\,:\,#2\big\}}

\newcommand{\II}[2]{\mbox{$\left[\,#1\,,\,#2\,\right]$}}
\newcommand{\Ii}[2]{\mbox{$\left[\,#1\,,\,#2\,\right)$}}

\ifthenelse{\equal{\numberingtheoremsectionyesno}{yes}}
{\newtheorem{lem}{Lemma}[section]}
{\newtheorem{lem}{Lemma}}
\newtheorem{defi}[lem]{Definition}
\newtheorem{theo}[lem]{Theorem}
\newtheorem{cor}[lem]{Corollary}
\newtheorem{rem}[lem]{Remark}
\newtheorem{pro}[lem]{Proposition}

\newcommand{\calD}{\mathcal{D}}
\newcommand{\calE}{\mathcal{E}}
\newcommand{\calH}{\mathcal{H}}

\newcommand{\calP}{\mathcal{P}}
\newcommand{\calX}{\mathcal{X}}

\newcommand{\calZ}{\mathcal{Z}}
\newcommand{\reals}{\mathbb{R}}

\newcommand{\map}[3]{#1\,:\,#2\rightarrow #3}
\newcommand{\mMap}[4]{#1\,:#2\ni#3\mapsto#4}
\newcommand{\Restr}[2]{#1 \! \mid _{#2}}

\newcommand{\p}{\partial}

\newcommand{\R}{\Rgen{}{}{}}

\newcommand{\D}{\Dgen{}{}{}}

\newcommand{\norm}[1]{|#1|}
\newcommand{\normlr}[1]{\left|#1\right|}
\newcommand{\Norm}[1]{\left|\!\left|#1\right|\!\right|}

\def\R{\mathbb R}

\def\N{\mathbb N}

\newcommand{\Normlo}[1]{\Norm{#1}_{1}}
\newcommand{\Normlt}[1]{\Norm{#1}_{2}}
\newcommand{\Normlp}[1]{\Norm{#1}_{p}}
\newcommand{\Normli}[1]{\Norm{#1}_{\infty}}
\newcommand{\pt}{\partial_t}
\newcommand{\px}{\partial_x}

\newcommand{\pp}{\partial_p}

\newcommand{\ptx}{\partial_{t, x}}

\newcommand{\pxp}{\partial_{x,p}}

\newcommand{\ptxp}{\partial_{t,x,p}}

\renewcommand{\rho}{\varrho}

\renewcommand{\D}{D^{[3]}}
\newcommand{\Drj}{\D_{\varrho j}}

\allowdisplaybreaks

\renewcommand{\extended}[1]{}
\title[Radiation Reaction]{\sf A non-relativistic Model of Plasma Physics Containing a Radiation Reaction Term}  
\author{Sebastian Bauer}
\address{Fakult\"at f\"ur Mathematik,
Universit\"at Duisburg-Essen, Campus Essen, Germany}
\email[Sebastian Bauer]{sebastian.bauer.seuberlich@uni-due.de} 
\date{March 2016}

\begin{document}

\begin{abstract}\noindent
While a fully relativistic collisionless plasma is modeled by the
Vlasov-Maxwell system  a good approximation in the  non-relativistic limit is
given by the Vlasov-Poisson system. We modify the Vlasov-Poisson
system so that damping due to the  relativistic effect of radiation
reaction is included. We prove the existence and uniqueness as well
as the higher regularity of local classical solutions. These theorems
also include the higher regularity of classical solutions of the Vlasov-Poisson system depending on the regularity of the initial
datum.
\end{abstract}

\maketitle

\setcounter{equation}{0}

\section {Introduction}

In studying interactions of charged matter and electromagnetic fields it is a quite common situation that the
velocity of  matter is small if compared to the velocity of light.
In such situations an asymptotic expansion of the equations in the small parameter $\frac{\norm{v}}{c}$ 
can help to reduce the complexity of the original problem, e.g. for numerical computations.
Since in many relativistic settings the zeroth order contribution of this expansion gives the Newtonian limit such expansions
are usually called post-Newtonian expansions.

In numerous applications of asymptotic analysis it is quite straightforward to establish
the equations of the expansion  and it is the main task to prove a relation between
the solutions of the asymptotic expansion and the solution of the original equation.
In contrast to that in interactions of charged matter and electrodynamic fields also
the formulation of the approximation scheme becomes delicate if effects due to radiation damping
shall be included.  It is well-known that accelerated charged matter loses energy by radiation; an electromagnetic 
radiation field is generated  which transports energy to null-infinity. This gives a damping 
effect on the motion of the matter known as radiation damping. There is a large
amount of literature for single particle models concerning effective equations which incorporate radiation 
damping without giving a fully relativistic description, see \cite{spohn:2004} and the literature cited therein. 

On the other hand  the Vlasov-Maxwell system is a well-known fully relativistic description
of a large ensemble of collisionless particles which interacts by means of the 
electromagnetic fields  collectively generated by these particles.
In a non-relativistic setting this situation is governed by the Vlasov-Poisson system.
While the asymptotic limit relation between the Vlasov-Poisson and the Vlasov-Maxwell system is well understood,
see \cite{schaeffer:86, lee:04}, and also higher order relations between the so called Vlasov-Darwin and the Vlasov-Maxwell
model are established, see \cite{bauerkunze:05}, all these approximation models do not inhibit radiation damping. 
 
It is the aim of this paper to establish and study a model in between the 
fully relativistic description and the non-relativistic description where
effects due to radiation damping are included. 
In Section \ref{Heuristics} we repeat some heuristics from \cite{kunzerendall:01b,kunzerendall:01a} 
leading to two candidates of such systems, see equations (\ref{2+1}) and (\ref{3+0}). While the first of these systems is already studied in 
\cite{kunzerendall:01b,kunzerendall:01a} we give an additional heuristic leading to (\ref{3+0}).
In the third section we shall reformulate (\ref{3+0}) establishing the topic of this
paper, the 'reduced Vlasov-Poisson system with radiation damping', see (\ref{rVPRD}).  

In this situation the main criterion to decide wether or not a model is suitable,  is the approximation property.
In a second paper \cite{bauer:16} it is proved that solutions of (\ref{rVPRD}) together with some lower order terms
approximate solutions of the Vlasov-Maxwell system 
to a higher order in $v/c$ than solutions of the Vlasov-Poisson 
or the Vlasov-Darwin model,
 see also \cite{bauerkunze:06} for an overview. 

In Section \ref{Proof} we present a proof of local existence, uniqueness 
and smoothness of solutions of (\ref{rVPRD}).
This proof is an adaption of  the usual proof of local existence of classical solutions, essentially going back to \cite{batt:77}.
In this contribution we follow the presentation in \cite{rein:07}.
With regard to the issue of higher regularity we have to use an additional induction loop.  

Motivated by a preprint version of this paper the existence of
global solutions for small initial data has already been proved in \cite{chenzhang:15}. 

\section{Heuristic derivations of models including radiation damping}\label{Heuristics}
\subsection{The Vlasov-Poisson system with radiation damping}
How could such a model look like? 
In \cite[Thm. 1.4]{bauerkunzereinrendall:06} it is shown that for suitable solutions of the Vlasov-Maxwell
system which are isolated from incoming radiation in the non-relativistic limit $c\to\infty$, $c$ the velocity of light,  the total
rate of radiated energy is given by
$$ \frac{2}{3c^3}\norm{\ddot{\mathcal D}(t)}^2$$
where $\mathcal D$ is the dipole moment of the non-relativistic charge distribution, i.e.~of the
dipole moment of the solution of the Vlasov-Poisson system. This result yields a mathematical 
formulation and rigorous proof of Larmor's formula, see \cite[(14.22)]{jackson:98}, in the case of Vlasov matter.

Therefore the goal is to implement an additional term into the Vlasov-Poisson system 
causing this loss of energy. As already suggested in \cite{kunzerendall:01b,kunzerendall:01a} we modify the Vlasov equation 
by incorporating a small additional force term  
\begin{equation}\label{Vlasov-equation}
	\pt f^\pm +p\cdot \nabla_x f^\pm \pm\left(E
    +\frac{2}{3c^3}\dddot{\mathcal D}\right) \cdot \nabla_p f^\pm =0\,.
\end{equation}
Here $f^+$ and $f^-$ give the distribution density of two species of charged particles
where the mass of the particles is set to unity and the charge of the particles is
$+1$ and $-1$, respectively. These densities are functions of time
$t\in \R$ and phase space variables $(x, p)\in \R^3\times \R^3$.
The resulting charge density is given by $\varrho$,
\begin{equation}\label{charge-density-def}
	\varrho^\pm(t, x)  =   \int  f^\pm(t, x, p)\,dp \quad\text{and}\quad \varrho=\varrho^+-\varrho^-\,,
\end{equation}
$E$ is the electrostatic
field generated by this charge density,
\begin{equation*}
	E(t, x)  =  -\nabla\int\frac{\varrho(t, y)}{|x-y|} \,dy\,,
\end{equation*}
and finally ${\mathcal D}$ is the dipole moment
of the charge distribution,
\begin{equation}\label{dipole-def}
	{\mathcal D}(t) =  \int x \varrho(t, x)\,dx\,.
\end{equation}
We will refer to this set of equations (\ref{Vlasov-equation})--(\ref{dipole-def}) as the 
{\sl Vlasov-Poisson system with radiation damping}.
The additional term in (\ref{Vlasov-equation}) is a generalization of the
radiation reaction force term used in particle models; see \cite[(16.8)]{jackson:98}.
We also note that for this system the quantity 
\begin{equation*}
  {\mathcal E}_{S}= \frac{1}{2} \iint
  p^2(f^++f^-)(t, x, p)\,dp\,dx 
  +\frac{1}{8\pi}\int |E(t,
  x)|^2\,dx-\frac{2}{3}c^{-3}\dot{\mathcal D}\cdot\ddot{\mathcal D} 
\end{equation*}
is decreasing, more precisely one obtains
\begin{equation*}
  \frac{d}{dt}{\mathcal E}_S(t) = -\frac{2}{3c^3}
  |\ddot{\mathcal D}(t)|^2, 
\end{equation*}
the subscript $S$ referring to the name ``Schott''-energy under which
this energy can be found in the literature. 	
This decreasing of energy
is usually attributed to the effect of radiation damping.
\extended{We give this computation in some detail:
		\begin{align*}
 			\frac{d}{dt} \calE_S(t) 
							&=\frac{1}{2}\iint p^2 \left(-p\cdot \px(f^++f^-)
								-(E+\frac{2}{3c^3}\dddot{\calD})\cdot \pp(f^+-f^-)\right)\,dp\,dx\\
							&\quad	+\frac{1}{4\pi}\int E\cdot \pt E\,dx-
									\frac{2}{3c^3}\norm{\ddot{\calD}}-
									\frac{2}{3c^3}\dot{\calD}\cdot\dddot{\calD}	\\
							& = 	\int \left(j^+-j^-\right)(E+\frac{2}{3c^3}\dddot{\calD})
									+\frac{1}{4\pi}\int E \cdot\px\Delta^{-1}\pt\varrho\,dx
							 		-\frac{2}{3c^3}\norm{\ddot{\calD}}^2-\frac{2}{3c^3}\dot{\calD}\cdot\dddot{\calD}\\
							& =  \int \left(j^+-j^-\right)(E+\frac{2}{3c^3}\dddot{\calD})
									-\int \varrho\Delta^{-1}\pt\varrho\,dx
							 		-\frac{2}{3c^3}\norm{\ddot{\calD}}^2-\frac{2}{3c^3}\dot{\calD}\cdot\dddot{\calD}\\
							& = 	\int \left(j^+-j^-\right)(E+\frac{2}{3c^3}\dddot{\calD})
									+\int (\Delta^{-1}\varrho)\,\px\cdot(j^+-j^-)\,dx
							 		-\frac{2}{3c^3}\norm{\ddot{\calD}}^2-\frac{2}{3c^3}\dot{\calD}\cdot\dddot{\calD}\\
							& = 	\int \left(j^+-j^-\right)(E+\frac{2}{3c^3}\dddot{\calD})
									-\int E\cdot (j^+-j^-)\,dx
							 		-\frac{2}{3c^3}\norm{\ddot{\calD}}^2-\frac{2}{3c^3}\dot{\calD}\cdot\dddot{\calD}\\
							& = 	\frac{2}{3c^3}\dddot{\calD}\int j^+-j^-\,dx
									-\frac{2}{3c^3}\norm{\ddot{\calD}}^2-\frac{2}{3c^3}\dot{\calD}\cdot\dddot{\calD}\\
							& = 	-\frac{2}{3c^3}\norm{\ddot{\calD}}^2
		\end{align*}
}   
\subsection{Problems with the Cauchy problem}
In order to formulate the Cauchy problem of this system we of course
have to give initial values for the particle densities $f^\pm$, but we
also have to fix data for the dipole moment. Let us for a moment assume
that we are equipped with a smooth solution of (\ref{Vlasov-equation})--(\ref{dipole-def}) such that
the spatial support of $f^\pm$ is compact on every time slice with
constant time. We define the bare mass by 
\begin{equation}\label{bare-mass-def}
M(t)  =  \iint\left(f^+ +f^-\right)(t, x, p)\,dp\,dx\,. 
\end{equation}
Then (\ref{Vlasov-equation}) yields mass conservation $\pt M=0$ as well as charge conservation  
$\pt \varrho^\pm+\nabla\cdot j^\pm=0$ for 
both species,  where 
\begin{equation}\label{current-density-def}
	j^\pm = \int p f^\pm\,dp
\end{equation} are the current densities.
Using equation (\ref{Vlasov-equation}) we compute the first three time derivatives of ${\mathcal D}$ and find
$\dot{\mathcal D}(t)  =  \int j^+-j^-\, dx$ as well as
 $$
  \ddot{\mathcal D}(t)  =  D^{[2]}(t)+\frac{2}{3c^3} M\,\dddot{\mathcal D}(t)
$$
where $D^{[2]}$ is defined by
\begin{eqnarray*}
   D^{[2]}(t) & = & \int E(t, x)\, \left(\varrho^+ +\varrho^-\right)(t,
   x)\,dx\,.
\end{eqnarray*}
\extended{The computation in detail:
\begin{align*}
  \ddot{\calD} 	&= \iint p\pt(f^+-f^-)\,dp\,dx\\
					&= \iint p \left(-p\cdot\px(f^+-f^-)-(E+2/(3c^3)\dddot{\calD})\cdot\pp(f^++f^-\right)\,dp\,dx\\
					&= \int E(\varrho^++\varrho^-)\,dx+\frac{2}{3c^3}M \dddot{\calD}
\end{align*}	}
Introducing the new variable 
\begin{equation*}
  y(t) := \ddot{\mathcal D}(t) \quad\text{and abbreviating}\quad \varepsilon:=\frac{2}{3c^3}M
\end{equation*}
we can recast (\ref{Vlasov-equation}) into 
\begin{equation}
  \mylabel{sgpp}
  \left.
    \begin{array}{rcl}
    \partial_t f^\pm & = & -p\cdot \nabla_x f^\pm \mp\Big(E +y
    -D^{[2]}\Big) \cdot \nabla_p f^\pm\\ 
      \varepsilon \dot{y} & = & y - D^{[2]}
    \end{array}\right\}\tag{SGPP${}_\varepsilon$}
\end{equation}
where we have to supply initial data $f^{\circ, \pm}$ and $y_0
\in \R^3$. While the physical setting only delivers initial data for the
particle distribution the question arises how to specify $y_0$ in
order to get a solution which is physically reasonable. 
We shall discuss now several approaches to deal with this problem.
\subsection{Methods of reduction}	The Vlasov-Poisson system with radiation damping serves as an
approximation of a relativistic system valid up to order $c^{-3}$ or $\varepsilon$.
On account of that it seems to be acceptable to modify this system by terms which are
formally of higher order in $\varepsilon$. In \cite{kunzerendall:01b} there are several ways of reduction
discussed, each of it  consisting of two basic steps in certain combinations: 
on the one hand a time derivative will be replaced modulo higher order terms via the Vlasov equation,
e.g. $\ddot{\mathcal D}=D^{[2]}+\mathcal O(\varepsilon)$ and on the other hand a time
derivative will be cancelled by a substitution in the characteristic equation $\mathcal X^{\pm} =\mathcal P^{\pm}$,
${\mathcal P}^{\pm}=\pm (E+\varepsilon\dddot{\mathcal D})$.
The so-called 2+1 reduction $\tilde{\mathcal X}^\pm={\mathcal X^\pm}$, $\tilde{\mathcal P}^{\pm}={\mathcal P}^{\pm}\mp\varepsilon\ddot{\mathcal D}$
and $\ddot{\mathcal D}\approx D^{[2]}$ leads to the Vlasov equation
\begin{equation}\label{2+1}
	\pt f^{\pm}+(p\pm\varepsilon D^{[2]})\cdot f^\pm \nabla_x\pm E\cdot\nabla_p f^{\pm}=0
\end{equation}
which is thoroughly studied in \cite{kunzerendall:01a}.
Using a 3+0 reduction scheme, replacing $\dddot{\mathcal D}$ by $\dot{D}^{[2]}$, we find the Vlasov-equation
\begin{equation}\label{3+0}
\pt f^\pm +p\cdot \nabla_x f^\pm \pm\left(E
    +\varepsilon\dot{ D}^{[2]}\right) \cdot \nabla_p f^\pm =0\,.
\end{equation}

\subsection{A reduction via geometric perturbation arguments}
Here we want to give an alternative formal derivation of equation (\ref{3+0}) 
using an analogy to single particle models. 
First note that in
case of a single particle the  problem of the physical reasonable initial value is discussed since the
beginning of the last century, see e.g. \cite[chapter 16]{jackson:98}.
In \cite{kunzespohn:01} it has been observed that in particle models
this problem has the structure of a geometric singular perturbation problem, and the 
"physical" dynamics are obtained on a center-like manifold of the full dynamics. 
For sake of discussion let us assume for a moment that the first
equation in (\ref{sgpp}) is an ODE instead of a PDE and to simplify notations
in this subsection $f$ denotes the
couple $(f^+, f^-)$. The same convention will be
used for $f^\circ=(f^{\circ, +}, f^{\circ,-})$. Then theorems from
singular geometric perturbation theory, see \cite{sakamoto:90}, support us
with invariant manifolds. More precisely, the manifold ${\mathcal
  M}_0=\{y=h_0(f)=D^{[2]}\}$ being invariant under the dynamics of
(SGPP${}_{\varepsilon=0}$) persists for small $\varepsilon>0$: For sufficiently
small $\varepsilon>0$ there is a  manifold ${\mathcal
  M}_\varepsilon=\{y=h_\varepsilon(f)\}$ being invariant under the dynamics of
(\ref{sgpp}) which remains  close to ${\mathcal M}_0$ in a suitable sense and
carries those trajectories which are bounded in $y$.  

Unfortunately, this theory does not apply in our case because we are
dealing with a phase-space of infinite dimension; thus the proof of the existence 
of an invariant manifold is hard. For sake of discussion we shall take the existence of a smooth
invariant manifold for granted and assume that it is given by means of
a smooth function $h_\varepsilon=h_\varepsilon(f^\circ)$, acting on
$C^\infty_0(\R^3\times\R^3)\times C^\infty_0(\R^3\times\R^3)$ and
taking values in $\R^3$.  The manifold ${\mathcal
  M_\varepsilon}=\{(f^{\circ}, h_\varepsilon(f^{\circ})\}$ is invariant under
the flow of (\ref{sgpp}) if the solution of (\ref{sgpp}) subject to
the initial conditions $(f^{\circ},y(0))= (f^{\circ},
h_\varepsilon(f^{\circ}))$ satisfies
\begin{equation}\label{heps-y}
	 y(t)=h_\varepsilon(f(t, \cdot, \cdot))
\end{equation}
for all times the solution exits. 
In order to find an appropriate approximation of (\ref{sgpp}) let us
assume that we are furnished with smooth maps $h_\varepsilon$ defining the
manifolds ${\mathcal M}_\varepsilon$ and that we are able to expand in $\varepsilon$, 
\begin{equation}\label{heps-exp}
	h_\varepsilon=h_0+\varepsilon h_1+{\mathcal O}(\varepsilon^2)\,.
\end{equation}
We plug (\ref{heps-y}) and (\ref{heps-exp}) into the second line of (\ref{sgpp}).
Assuming that we are allowed to exchange the order of differentiation with respect to  $t$ and 
			expansion in $\varepsilon$	we compute
\begin{equation*}
 	\varepsilon\dot{h}_0(f)+{\mathcal O}(\varepsilon^2)
		= h_0(f)-D^{[2]}+\varepsilon h_1(f)+{\mathcal O}(\varepsilon^2)\,.
\end{equation*}Equating powers of $\varepsilon$ we obtain
\begin{equation*}
 		h_0(f)=D^{[2]}=\int E(\varrho^++\varrho^-)\,dx\quad\text{and}\quad h_1(f)=\dot{D}^{[2]}=\frac{d}{dt}\int E(\varrho^++\varrho^-)\,dx\,.
\end{equation*} 
Note that $D^{[2]}$ is a
function of $E$ and $\varrho^\pm$. But because $E$ itself is defined by means
of the Poisson integral with source $\varrho$ it is a function of
$f^\pm$, too. Therefore $D^{[2]}=D^{[2]}(f)$. 
Neglecting higher order terms we also end with the Vlasov-equation (\ref{3+0}), which shall be discussed in the following.\footnote{On a very formal level
 $h_{\varepsilon}(f)$ is given by $\sum_{j=0}^\infty\varepsilon^j D^{[2],(j)}$, where, using the Vlasov equation, all derivative w.r.t to 
time can be replaced by derivatives  w.r.t to phase-space.}
\section{The reduced Vlasov-Poisson system with radiation damping, main results}
\subsection{Reformulation using principal values}
First  we have to remove an obstacle. On the one hand we need  $\pt f(t)$ and  $f(t)$ 
to define $h_1$; therefore $h_1$ is not defined on 
$C^\infty_0(\R^3\times\R^3)\times C^\infty_0(\R^3\times\R^3)$. 
On the other hand in \cite[p. 3579]{kunzerendall:01b} it is argued 
that (\ref{3+0}) is a borderline case for the usual proof of local existence of classical solutions:
In order to obtain a local existence theorem the orders of differentiability of the different unknowns 
should fit  together properly. If it is supposed that the potential $u$ of $E$ is $k$-times differentiable,
then $f$ and $\varrho$ are $k-2$ times differentiable and thus by solving the Poisson equation 
$\Delta U = 4\pi \varrho$ two orders of differentiability should be gained which is not the 
case in spaces of pointwise differentiable functions.

We shall get rid of both of these problems by using the Vlasov-equation (\ref{3+0}) and integration by parts one more time,
but know we are forced to use principal values.
To this end we additionally introduce the notation $E^\pm=-\nabla\int  \varrho^\pm(y)\,\frac{dy}{|x-y|}$; 
then some elementary calculations,
using the identity $\nabla_x\norm{y-x}^{-1}=-\nabla_y\norm{y-x}^{-1}$, lead to 
\begin{align*}
 h_1(f) &= -\frac{d}{dt}\iint \nabla_x\left(\norm{x-y}^{-1}\right)(\varrho^+(y)-\varrho^-(y))(\varrho^+(x)+\varrho^+(x))\,dy\,dx \\	&=2\int E^+\,\partial_t\varrho^- -E^-\partial_t
\varrho^+\,dx.
\end{align*}
\extended{In full detail:
	\begin{align*}
 		h_1(f) &= -\frac{d}{dt}\iint \nabla_x\left(\norm{x-y}^{-1}\right)(\varrho^+(y)-\varrho^-(y))(\varrho^+(x)+\varrho^+(x))\,dy\,dx \\
		&= -\iint \nabla_x\left(\norm{x-y}^{-1}\right)\left\{(\pt \varrho^+(y)-\pt \varrho^-(y))(\varrho^+(x)+\varrho^-(x))
				+(\varrho^+(y)-\varrho^-(y))	(\pt\varrho^+(x)+\pt\varrho^-(x))\right\}	dy\,dx\\
		&= +\iint \nabla_y\left(\norm{x-y}^{-1}\right)(\pt \varrho^+(y)-\pt \varrho^-(y))(\varrho^+(x)+\varrho^-(x))\,dx\,dy\\
		&\quad -\iint \nabla_x\left(\norm{x-y}^{-1}\right)(\varrho^+(y)-\varrho^-(y))	(\pt\varrho^+(x)+\pt\varrho^-(x))	dy\,dx\\
		& =-\int (E^+(y)+E^-(y))(\pt \varrho^+(y)-\pt \varrho^-(y))\,dy\\
		&\quad+\int (E^+(x)-E^-(x))(\pt\varrho^+(x)+\pt\varrho^-(x))\,dx\\
		&= 2\int E^+(x)\pt \varrho^-(x)-E^-(x)\pt \varrho^+(x)\,dx	
\end{align*}}
Using charge conservation 
$ \partial_t\varrho^\pm+\nabla\cdot j^\pm = 0$
we e.g. compute
\begin{align*}
 	\int  E^+\,\partial_t\varrho^-\,dx 
	&= -\int E^+ \nabla\cdot j^-\,dx
	=-\iint\frac{x-y}{\norm{x-y}^3}\varrho^+(y)\nabla_x\cdot j^-(x)\,dy\,dx\\	
	&= -\int\varrho^+(y)\int \frac{z}{\norm{z}^3}\nabla_z\cdot j^-(y+z)\,dz\,dy\\
	&=-\int \varrho^+(y) \left(\lim_{\eta\to 0}\int_{\norm{z}>\eta} \frac{z}{\norm{z}^3}\nabla_z\cdot j^-(y+z)\,dz\right)\,dy\\
	&=-\int \varrho^+(y)\lim_{\eta\to 0}\left(-
	\int_{\norm{z}>\eta}\norm{z}^{-3}\left(-3{z\otimes z}{|z|^{-2}}+{\rm id}\right) j^-(y+z)\,dz\right.\\
	&\qquad\left.
	-\int_{\norm{z}=\eta}{z\oplus z}{\norm{z}^{-4}}j^-(y+z)\,dS(z)\right)\,dy\\
	&= \int \varrho^+H(j^-)\,dy\,.
\end{align*}
Here ${\rm id}$ denotes the identity mapping in $\R^3$ and for a suitable function $j$ the operator $H$ is defined by 
\begin{equation}
  \mylabel{H-def}
  H(j)(y):=\oint \frac{1}{|z|^3}\Big(-3\frac{z\otimes z}{|z|^2}+{\rm
  id}\Big)j(y+z)\,dz +\frac{4\pi}{3}j(y)\,.
\end{equation}
Note that the kernel
$ K(z)=-3\frac{z\otimes z}{|z|^2}+{\rm id}$
is bounded on $\R^3\setminus\{0\}$, is homogeneous of
degree zero and satisfies
$\int_{|z|=1}K(z)\,d\sigma(z)=0$.
Thus, using the Calder{\'o}n-Zygmund inequality, $H$ is well defined for
smooth functions with compact support and for $1<p<\infty$
can be extended to a bounded linear operator mapping $L^p(\R^3)$ to $ L^p(\R^3)$, see \cite{stein:70}.
\extended{With regard to the boundary integral note that $z/\norm{z}$ is the {\sl inner} normal. Furthermore
$$\int_{\norm{z}=\eta}z_iz_j\, do(z)=0$$
if $i\neq j$ and
$$ \int_{\norm{z}=\eta}z_i^2\, do(z)=\frac{1}{3}\int_{\norm{z}=\eta}\norm{z}^2\, do(z)=\frac{4\pi}{3}\eta^4\,.$$}
We  call the following set of equations the  "reduced Vlasov-Poisson system with radiation damping"
\begin{equation}\label{rVPRD}\tag{${\rm rVPRD}_{\varepsilon}$}
		\begin{split}
			\lefteqn{\pt f^\pm+p\cdot\nabla_x f^\pm \pm (E +\varepsilon\D)\cdot\nabla_p f^\pm=0\,,}\\
			E(t, x) & =  -\nabla_x\displaystyle\int \,\frac{\varrho(t, y)}{|x-y|}\,dy\,,\\
			\D(t) & =  \displaystyle 2 \int\Big(H(j^-)\,\varrho^+
   -H(j^+)\,\varrho^-\Big)\,(t, x)\,dx\,,
		\end{split}
\end{equation}
where $\varrho^\pm$, $j^\pm$ and $H(j^\pm)$ are defined according to (\ref{charge-density-def}), (\ref{current-density-def})
and (\ref{H-def}).
To be precise we shall define what we mean by a classical solution.
We adopt the formulation of the usual Vlasov-Poisson system given in \cite{rein:07}. 
\begin{defi}\label{solution-def}
The pair of functions $\map{f^\pm}{J\times\R^3\times\R^3}{\Ii{0}{\infty}}$ is a classical solution 
of the reduced Vlasov-Poisson system with radiation damping on the interval $J\subset\R$ if the following holds:
\begin{itemize}
	\item[(i)] 	The functions $f^\pm$ are continuously differentiable with respect to all its variables.
	\item[(ii)]	The induced spatial densities $\varrho^\pm$ and $j^\pm$, the induced electric field $E$ and
				the induced damping term $\D$ exist on $J\times\R^3$ and $J$, respectively, and are continuously 
				differentiable.
	\item[(iii)] For every compact subinterval $I\subset J$ the electric field $E$ is bounded on $I\times\R^3$.
	\item[(iv)]	 The functions $f^\pm$,  $E$ and $\D$ satisfy the equations in (\ref{rVPRD}). 
\end{itemize}
\end{defi}

\subsection{Local solutions, uniqueness and higer regularity}
It is the main result of this paper to establish local existence and uniqueness  of
classical solutions of this system. Furthermore we want to establish
higher regularity of solutions of (\ref{rVPRD}) depending on the
smoothness of the initial data. To the best knowledge also in the case
$\varepsilon=0$, which is he usual Vlasov-Poisson system, this question is only
investigated in \cite{lindner:1991} and not published elsewhere. Surprisingly
in that paper only smoothness in spatial directions is proved whereas
the methods also allow for proving smoothness for all space-time
directions. At last we want to establish certain bounds holding
uniformly in $0\le\varepsilon\le 1$. These bounds are needed in \cite{bauer:16}
in order to prove the asymptotic approximation of solutions of the Vlasov-Maxwell
systems by densities which are built from solution of 
(\ref{rVPRD}).\footnote{For the approximation we also need some correction terms of 'Darwin-order'}
We shall take our initial data from the following class of smooth
functions with compact support. Fix an integer $k$ and constants $R_0$
and $S_0$. With regard to the initial data we require that
\begin{equation}\mylabel{bas-con-def}
	\begin{array}{c}
		f^{\circ, \pm} \in C^\infty_0(\R^3\times\R^3)\,,\qquad
		\Norm{ f^{\circ, \pm}}_{W^{\infty, k} (\R^3\times\R^3)} \le S_0\,,\\[0.5em] 
		f^{\circ, \pm} (x, p) = 0\qquad \text{if} \quad |x|\ge R_0 \quad
   		\text{or} \quad |p|\ge R_0 \,.
\end{array}
\end{equation} 
Here $\Norm{ \cdot}_{W^{\infty, k} (\R^3\times\R^3)}$ denotes the usual Sobolev-norm. For further notations also in the following 
theorem, see the beginning of Section \ref{Proof}.
Now we can state the main result of this paper.
\begin{theo}
  \mylabel{main-th}
   Fix a positive integer $k$, constants $R_0$, $S_0$ and initial data
   $f^{\circ, \pm}$ according to (\ref{bas-con-def}). Then for every
   $0\le \varepsilon\le 1$ the problem (\ref{rVPRD}) has a unique classical
   solution
   \[ \Big(f^\pm, E, \D \Big)\in C^k([0, T_\varepsilon)\times \R^3\times
   \R^3)\times C^k([0, T_\varepsilon)\times\R^3)\times C^k([0, T_\varepsilon)) \]
with $0< T_\varepsilon\le \infty$.
Furthermore, there is a constant $\hat{T}\leq \infty$ and for every constant
$0<T<\hat{T}$ there is a constant $M(T)$ such that  for
  all $x\in\R^3$, $p\in \R^3$ and $0\le t\le T$
\begin{itemize}
\item[(a)] $\hat{T}\le T_\varepsilon$ for all $0\le\varepsilon\le 1$,
\item[(b)] $f^\pm(t, x, p)=0$ if $|x|\ge M(T)$ or $|p|\ge M(T)$
\item[(c)] $\Normli{\ptxp^l f^\pm(t)}+\Normli{\ptx^l E(t)}+|\frac{d^l}{dt^l} \D (t)|\le M(T)$ for all 
 non-negative integers $l\le k$.
\end{itemize}
In addition
\begin{itemize}
  \item[(d)] $\dot{D}{}^{[2]}(t) = \D (t)$ for all $0\le t \le
  T_\varepsilon$. 
\end{itemize}
The constants $\hat{T}$ and $M(T)$ are only depending on $k$, $R_0$
and $S_0$. In 
particular they are independent of $\varepsilon$. 
\end{theo}

The proof is elaborated in Section \ref{Proof}. 
\subsection{Global solutions}
Next we turn to the question of existence of global in time classical solutions.
Motivated by a preprint version of this paper  global solvability for small initial
data has already been proved in \cite{chenzhang:15}. 

One crucial ingredient of every proof of existence of global solution with unrestricted initial data is an a-priori bound on the 
second velocity momenta $\iint p^2 f^\pm\,dp\,dx$.\footnote{Very recently in \cite{chenzhang:16} a proof of the existence of global classical solutions of the Vlasov-Poisson system has been published, which does not rely on an a-priori bound of the second velocity momentum. It remains to investigate wether this proof can be applied to the system at hand.} This bound is usually obtained by some energy 
conservation equation.  We introduce the  notation
$$ D^{[1]}:=\int j^+-j^-\,dx =\dot{\mathcal D}$$ and compute $\dot{D}^{[1]}= D^{[2]}+\varepsilon M D^{[3]}$ for a solution of (\ref{rVPRD}), 
 where $M$ is the time-independent bare mass, see (\ref{bare-mass-def}). 
We recall $\dot{D}^{[2]}= D^{[3]}$ 
and define the energy of the system (\ref{rVPRD}) by 
\begin{equation}
  \label{StEne-def}
  \tilde{\mathcal E}_{S}= \frac{1}{2} \iint
  p^2(f^++f^-)\,dp\,dx 
  +\frac{1}{8\pi}\int |E|^2\,dx-\varepsilon D^{[1]} \cdot D^{[2]}+\frac{\varepsilon^2M}{2}\norm{D^{[2]}}^2\,. 
\end{equation}
This energy is decreasing, more precisely we have
\begin{eqnarray*}
  \frac{d}{dt}\tilde{\mathcal E}_S & = & 
 -\varepsilon |{D^{[2]}}(t)|^2\,.
\end{eqnarray*}
\extended{The  computation is the same as in the case of Schott energy:
		\begin{align*}
 			\frac{d}{dt} \calE_S(t) 
							&=\frac{1}{2}\iint p^2 \left(-p\cdot \px(f^+-f^-)
								-(E+\varepsilon\D)\cdot \pp(f^++f^-)\right)\,dp\,dx\\
							&\quad	+\frac{1}{4\pi}\int E\cdot \pt E\,dx-
									\varepsilon\norm{D^{[2]}}^2-
									\varepsilon D^{[1]}\cdot\D	\\
							& = 	\int \left(j^+-j^-\right)(E+\varepsilon\D)
									+\frac{1}{4\pi}\int E \cdot\px\Delta^{-1}\pt\varrho\,dx
							 		-\varepsilon\norm{D^{[2]}}^2-
									\varepsilon D^{[1]}\cdot\D\\
							& =  \int \left(j^+-j^-\right)(E+\varepsilon\D)
									-\int \varrho\Delta^{-1}\pt\varrho\,dx-
							 		\varepsilon\norm{D^{[2]}}^2-
									\varepsilon D^{[1]}\cdot\D\\
							& = 	\int \left(j^+-j^-\right)(E+\varepsilon\D)
									+\int (\Delta^{-1}\varrho)\,\px\cdot(j^+-j^-)\,dx
							 		-\varepsilon\norm{D^{[2]}}^2-
									\varepsilon D^{[1]}\cdot\D\\
							& = 	\int \left(j^+-j^-\right)(E+\varepsilon\D)
									-\int E\cdot (j^+-j^-)\,dx-
							 		\varepsilon\norm{D^{[2]}}^2-
									\varepsilon D^{[1]}\cdot\D\\
							& = 	\varepsilon\D\int j^+-j^-\,dx-
									\varepsilon\norm{D^{[2]}}^2-
									\varepsilon D^{[1]}\cdot\D\\
							& = 	-\varepsilon\norm{D^{[2]}}^2
		\end{align*}
}   
Obviously, this energy is indefinite. This situation is well known from the Vlasov-Poisson equation in the gravitational case,
where the potential energy is negative. Here the question is, wether or not it is possible to estimate $\norm{D^{[1]}\cdot D^{[2]}}$ 
against one of the immanently positive terms in a suitable way. 
Of course we can estimate 
$2\norm{D^{[1]}\cdot D^{[2]}}\le \frac{\norm{D^{[1]}}^2}{\eta}+\eta\norm{D^{[2]}}^2$ for all $\eta>0$; therefore, it would be 
sufficient to bound $\norm{D^{[1]}}^2$. Unfortunately, using  the usual bounds for velocity momenta, see e.g. 
\cite[Lemma 1.8]{rein:07}, 
we only get the estimate
\begin{equation}\label{M1-est}
	\Normlo{j^\pm}\le C \Normlo{f^\pm}^{1/2} \left(\iint p^2 f^\pm\,dp\,dx\right)^{1/2}\,,
\end{equation}
which is just not sufficient for our purposes.

We also can't make use of the smallness of $\varepsilon$; in order to swallow the $\norm{D^{[2]}}^2$ term we have to choose
$\eta=\frac{\varepsilon M}{2}$ and end with the factor $\frac{1}{M}$ in front of the $\norm{D^{[1]}}^2$ term which is
just cancelled by the square of $\Normlo{f^{\pm}}^{1/2}$ in (\ref{M1-est}).
It is because of this, that the existence of global classical solutions seems to be out of reach at the moment.
If we take the boundedness of the second velocity momenta for granted
 there are no obvious obstacles to prove the existence 
of global solutions following the lines of the proof of Lions and Perthame in \cite{lionsperthame:91}.\footnote{
Note that for a solution of (\ref{2+1}) the usual total energy
${\mathcal E}= \frac{1}{2} \int\!\!\!\int p^2(f^++f^-)\,dp\,dx 
  +\frac{1}{8\pi}\int |E|^2\,dx$ is decreasing, $\pt {\mathcal E}(t) = -\varepsilon \norm{D^{[2]}}^2$. 
Using the resulting a-priori bound on 
$\int\!\!\!\int p^2(f^++f^-)\,dp\,dx$ global existence of smooth solutions has been shown in \cite{kunzerendall:01a}.}  

\section{Proof of Theorem \ref{main-th}}\label{Proof}
Theorem \ref{main-th} will be proved in two steps. After collecting some well known estimates in Subsection \ref{prep} 
we define a modification of the usual iteration scheme in Subsection \ref{iterates} and establish certain bounds on the iterates. 
In Theorem \ref{C1solution-theo} in subsection \ref{C1solution}, 
we prove that the iterates converge to a solution of (\ref{rVPRD}) and that this solution is unique. 
Finally,  in Subsection \ref{Sec-k} we shall prove 
higher regularity  of solutions, see Theorem \ref{Theo-k}.
Together Theorem \ref{C1solution-theo} and Theorem \ref{Theo-k} give Theorem \ref{main-th}.

Throughout this section we fix a positive integer $k$ and some
constants $R_0$ and $S_0$.  Generic constants which may be computed
from $k, S_0$ and $R_0$ are denoted by $C, C_1, C_2$ and so on.  Constants $C$ may change from line to
line. 
For $1\le p \le \infty$ we denote the spatial-space and the phase-space $L^p$ norms by $\Normlp{\cdot}$.
If we consider  a pair of functions then  $\Normlp{\cdot}$ gives the sum of the norms of the pair, e.g.
$\Normlp{f^\pm}=\Normlp{f^+}+\Normlp{f^-}$.
For $t\in\R$ we denote by $f(t)$ the function
$$ \mMap{f(t)}{\R^3\times\R^3}{(x,p)}{f(t,x,p)}$$
and in the same sense we use $\varrho(t)$, $j(t)$ and $E(t)$. By $C(\R^n)$ and $C^k_c(\R^n)$ we denote the space 
of $k$ times continuously differentiable functions on $\R^n$, the subscript $c$ indicates compactly supported functions.
For differentiable functions $f$, we denote (the vector of) partial derivatives with respect to $t$, $x$, $p$ or combinations of it by   
$\pt f$, $\px f$, $\pp f$, $\pxp f$ and so on. Vectors of partial derivatives of order $l$ will be denoted by
$\pxp^l f$, $\ptxp^l f$ and so on. The subscripts indicate which partial derivatives are included.

\subsection{Preparations}\label{prep}

For $\varrho\in C_c(\R^3)$ we define
	$$ E_\varrho(x) = -\nabla_x\int\frac{\varrho(y)}{|x-y|}\, dy =\int\frac{x-y}{|x-y|^3}\varrho(y)\,dy\,. $$
Furthermore we set $\ln^*(a):=1+\ln(a)$ if $1\le a$ and $\ln^*(a)=a$ if $0\le a <1$.
For the convenience of the reader we cite the following propositions from
\cite[Prop.~1, Prop.~2]{batt:77}.

 \begin{pro}\mylabel{Batt}
    Let $\varrho\in C_c^1(\R^3)$. Then $E_\varrho\in C^1(\R^3)$ and  for all $0<d\le R$
\begin{align}
\tag{i}	\Normli{E_\varrho} &\le 3(2\pi)^{2/3}
\Normlo{\varrho}^{1/3}\Normli{\varrho}^{2/3}\,,\\
\tag{ii}	\Normli{\px E_\varrho} &\le C \left[R^{-3}\Normlo{\varrho}+d\Normli{\px \varrho}+\left(1+\ln \left(R/d\right)\right)\Normli{\varrho}\right]\,,\\
\tag{iii}\label{dxE2}	\Normli{\px E_\varrho} &\le C \left[\big(1+\Normli{\varrho}\big)\big(1+\ln^*\Normli{\px \varrho}\big)+\Normlo{\varrho}\right]\,.
\end{align}
  \end{pro}

For $\varrho\in C_c(\R^3)$ and $j\in C_c(\R^3;\R^3)$ we define
$$ \Drj:= 2\int \varrho(x) \left[H(j)\right](x)\,dx\,.$$

\begin{lem}\label{Calderon}
	Let $\varrho\in C_c(\R^3)$ and $j\in C_c(\R^3;\R^3)$. Then
		$$\norm{\Drj}\le 2\ccz\Normlt{\varrho}\Normlt{j} 
			\le  2\ccz \big(\Normli{\varrho}\Normlo{\varrho}\Normli{j}\Normlo{j}\big)^{1/2}\,,$$
	where $\ccz$ is the Calder{\'o}n-Zygmund constant.
	Let $J\subset\reals$ an intervall, $\varrho\in C(J;C_c(\R^3))$ as well as $j\in C(J;C_c(\times\R^3;\R^3))$.
	Then the function $J\ni t\mapsto \Drj(t):=\D_{\varrho(t)j(t)}$ is continuous. If additionally $\varrho$ and $j$ are continuously differentiable
	with respect to  $t$, then $\Drj$ is differentiable and the usual product rule holds true.
	
\end{lem}
\begin{proof}
Applying H\"older's inequality, the Calder{\'o}n-Zygmund inequality \cite[pp.39,Thm.3]{stein:70},  and 
H\"older's inequality again
proves the estimate. The latter statements follow by standard theorems on Lebesgue-integrable functions. 
\end{proof}

For an intervall $J\subset \R$ and a mapping $F\in C(J\times\R^3;\R^3)$, $ t\in J$ and $z=(x,p)\in\R^3\times\R^3$ we define
${\mathcal Z}_F=({\mathcal X}_F, {\mathcal P}_F)=({\mathcal X}_F, {\mathcal P}_F)(s; t , x,
  p)$ as the unique solution of 
\begin{equation*}
   \begin{array}{rclcrcl}
   \displaystyle\frac{d}{ds} {\mathcal X}_F(s) & = &\,{\mathcal P}_F(s), &\qquad&
    {\mathcal X}_F(t; t, x, p)
    & = & x\,,\\[2ex] 
    \displaystyle\frac{d}{ds} {\mathcal P}_F(s) & = & \,F(s, {\mathcal X}_F(s)), &\qquad& {\mathcal P}_F(t; t,
    x, p) &  = & p\,.
   \end{array}
\end{equation*}
Furthermore, we set
\begin{align*}
	P_{\calZ}(t) 	&:=\sup\setb{\norm{\calP_F(s;0,\tilde{x},\tilde{p})}}{0\le s\le t,\, \norm{\tilde{x}}\le R_0,\,\norm{\tilde{p}}\le R_0}\,, \qquad X_{\calZ}(t) :=R_0+\int^t_0P_{\calZ}(s)\,ds\\ 
	C_{\calZ}(t)	 	& :=\sup\set{\norm{\pxp\calZ_F(s;t,x,p)}}{0\le s\le t,\,  x,p\in\R^3}
\end{align*}
For  an  initial datum $f^\circ$ according 
to (\ref{bas-con-def}) we define
\begin{align}\nonumber
	f_F(t, x, p)&:=f^\circ(\calZ_F(0;t,x,p)).
\end{align}
We collect some well known facts about the flow generated by a smooth bounded field. 
For a proof see e.g. \cite[Lemma 1.2, Lemma 1.3]{rein:07}.
\begin{lem}\label{flow}
	Assume that $F$ is additionally continuously differentiable with respect to $x$ and bounded on 
	$I\times\R^3$ for every compact interval $I\subset J$. Then
	\begin{itemize}
		\item[(i)] 	$\calZ_F\in C^1(J\times J\times\R^3\times \R^3)$ and for all $(s,t)\in J\times J$  the map
					$\calZ_F(s;t,\cdot,\cdot)$ is a measure preserving $C^1$-diffeomorphism with inverse
					$\calZ_F(s;t,\cdot,\cdot)^{-1}=\calZ_F(t;s,\cdot,\cdot)$.
		\item[(ii)]	$P_{\calZ}(t) \le \diss R_0+\int^t_0\Normli{F(s)}\,ds$ for all $t\in J.$
		\item[(iii)]  $\diss C_{\calZ}(t)\le 6\exp \left(\int_0^t1+\Normli{\px F(s)}\,ds\right)$ 
						for all $t\in J$.
		\item[(iv)] $f_F\in C^1(J\times\R^3\times\R^3)$, $f_F(t,x,p) =0 $ if $\norm{x}> X_{\calZ}(t)$ or 
					$\norm{p}> P_{\calZ}(t)$, $\Normli{f_F(t)}=\Normli{f^\circ}$  and $\Normlo{f_F(t)}=\Normlo{f^\circ}$
					for all $t\in J$.
		\item[(v)] $f_F$ is the unique $C^1$-solution of the Vlasov equation
					$$\pt f+p\cdot\px f+F\cdot\pp f =0 \quad\text{with}\quad f(0)=f^{\circ}\,.$$
	\end{itemize}	 
\end{lem}
\extended{
\begin{proof} The proof of these statements is well known and only repeated for the convenience of the reader.
	According to the smoothness and growth assumptions of $F$ 
	existence and smoothness of $\calZ_F$ and (ii) follow by standard
	theorems of ODE. Furthermore, because of $\pxp\cdot (p, F(t, x))^t=0$
	Liouville's Theorem implies that the generated flow $\calZ$ is volume 
	preserving. 
	With regard to (iii) we note that $\pxp\calZ_F(\cdot;t,x,p)$ is the solution of the initial problem
\begin{equation*}
   \begin{array}{rclcrcl}
   \displaystyle\frac{d}{ds} \pxp\calX_F(s) & = &\pxp\calP_F(s), &\qquad&
    {\pxp\calX}_F(t; t, x, p)
    & = & \pxp x\\[2ex] 
    \displaystyle\frac{d}{ds} \pxp\calP_F(s) & = & \,(\pxp F)(s,\calX_F)\cdot\pxp\calX_F, &\qquad&\pxp\calP_F(t; t,
    x, p) &  = &\pxp p\,,
   \end{array}
	\end{equation*}
	which is a homogenous linear ODE system. Thus, applying  a standard estimate of ODE theory 
	the proposed inequality follows.
	By definition of $f_F$ and since $\calZ_F$ is measure preserving the equalities in (iii) follow.
	Now take $p\in\R^3$ with $\norm{p}>P_\calZ(t)$. Take some $x\in\R^3$ and define
	$(\tilde{x},\tilde{p}) = \calZ_F(0;t,x,p)$. Since $\norm{p}>P_F(t)$ we have $\norm{\tilde{p}}> R_0$ and/or 
	$\norm{\tilde{x}}>R_0$. Anyway we conclude $f_F(t, x,p)=f^\circ (\tilde{x},\tilde{p})=0$. 
	Now take some $x\in\R^3$ with $\norm{x}> X_F(t)$: Pick any $p\in\R^3$ and define again
	$(\tilde{x},\tilde{p}) = \calZ_F(0;t,x,p)$. If we assume that $\norm{\tilde{x}}\le R_0$ and $\norm{\tilde{p}}\le R_0$
	we conclude
		\begin{align*}
 			\norm{x} 	&\le \norm{\tilde{x}+\int_0^s\calP_F(\sigma;0,\tilde{x},\tilde{p})\,d \sigma }\\
						&\le R_0+\int_0^t P_F(\sigma)\, d \sigma=X_f(t)\,.
			\end{align*}
	Therefore, we conclude $\norm{\tilde{p}}> R_0$ and/or 
	$\norm{\tilde{x}}>R_0$ and $f_F(t, x, p)=f^\circ (\tilde{x},\tilde{p})=0$.	
\end{proof}}
\subsection{The iteration scheme and estimates on the iterates}\label{iterates}
We set up a modification of the usual iteration scheme.  Fix initial
data $f^{\circ, \pm}$ according to (\ref{bas-con-def}).
For $n=0$, $s,t\in [0,\infty)$ and $x,p\in\R^3$ we define
$ \calZ^\pm_0(s;t,x,p)=(x,p)$.
If $\calZ^\pm_n$ is defined we set 
\begin{align*}
	f^\pm_n(t, x, p) & =  f^{\circ, \pm}(\calZ_n^\pm(0;t,x,p)) & & \\
	P_{n}(t)	 		& = \max \left\{P_{\calZ^+_n}(t), P_{\calZ^-_n}(t) \right\}&
	X_{n}(t)	 		& = R_0+	\int_0^tP_{n}(s)\, ds\\
	\rho^\pm_n(t,x)& =  \int f^\pm_n(t,x,p)\,dp &
  	j^\pm_n(t,x)    	& =  \int pf^\pm_n(t,x,p)\,dp\\
	E_n    				& =  E_{\varrho_n^+} -E_{\varrho_n^-}&
	\D _n 				& =  \D_ {\varrho^+_n j^-_n}-\D_{\varrho^-_n j^+_n} \\
	F_n 				& = E_n+\varepsilon \D_n&
	\calZ^\pm_{n+1} & = \calZ_{\pm F_n}\,.
\end{align*}

\begin{lem}\label{Iteration}
	The iteration scheme is well defined and for all $t\ge 0$, $n\in\N$ and $x,p\in\R^3$ and $\varepsilon\in\II{0}{1}$ we have
	\begin{align}\label{f_n}
 		\tag{i} &f_n^\pm \in C^k([0,\infty)\times\R^3\times\R^3)\,, &
				f^\pm_n(t,x,p) &=0\quad\text{if}\quad\norm{x}\ge X_n(t) \quad\text{or}\quad\norm{p}\ge P_n(t)\,,\\
		\nonumber
			&\Normli{f^\pm_n(t) }=\Normli{f^{\circ,\pm}}\,, & \Normlo{f^\pm_n(t)} &=\Normlo{f^{\circ,\pm}}\,,\\
		\tag{ii}\label{rho_n}
				&\varrho_n^\pm \in C^k([0,\infty)\times\R^3)\,, & \varrho^\pm_n(t,x) &=0\quad\text{if}\quad\norm{x}\ge X_n(t)\,, \\
		\nonumber
				&\Normli{\varrho^\pm_n(t)}\le P_n^3(t)\Normli{f^{\circ,\pm}}	\,,&\Normlo{\varrho^\pm_n(t)}&=\Normlo{f^{\circ,\pm}}\,,\\
		\tag{iii}\label{jn}
			& j_n^\pm\in C^k([0,\infty)\times\R^3;\R^3)\,, &j^\pm_n(t,x)&=0\quad\text{if}\quad\norm{x}\ge X_n(t)\,, \\
		\nonumber
			&\Normli{j^\pm_n(t)}\le P_n^4(t)\Normli{f^{\circ,\pm}} \,,& \Normlo{j^\pm_n(t)}&\le P_n(t)\Normlo{f^{\circ,\pm}}\,,\\
		\tag{iv}\label{E_n}
				&E_n\in C^k([0,\infty)\times\R^3;\R^3)\,,&\Normli{E_n(t)}&\le C_1P_n^2(t)
				\quad\text{with}\quad C_1=3 (2\pi)^{3/2}\Normli{f^{\circ,\pm}}^{2/3}\Normlo{f^{\circ,\pm}}^{1/3}\,,\\
		\tag{v}\label{D_n}
				&\D_n\in C^k([0,\infty);\R^3)\,,
				&\norm{\D_n(t)}&\le C_2P_n^4(t)\quad\text{with}
					\quad C_2=2\ccz\Normli{f^{\circ,\pm}}\Normlo{f^{\circ,\pm}}\,.
	\end{align}
\end{lem}
\begin{proof}
Using Proposition \ref{Batt}, Lemma \ref{Calderon} and Lemma \ref{flow} this Lemma follows immediately by induction on $n\in\N$.
\end{proof}

Next we shall give an 
estimate of the size of the support uniformly in $n$. 
Let $\map{P}{[0,a)}{\R}$ denote the maximal solution of the equation
\begin{equation}\label{P_def}
	P(t)=R_0+C_3\int_0^tP^2(s)+P^4(s)\,ds \quad\text{where}\quad C_3:=\max \left\{C_1, C_2\right\}\,,
\end{equation}
and define 
$$X(t):=R_0+	\int_0^tP(s)\,ds\,.$$

\begin{lem}\mylabel{P-Lemma}
	For all $0\le t<a$, $n\in\N$ and $\varepsilon\in[0,\,1]$ we have $P_n(t)\le P(t)$ and $X_n(t)\le X(t)$	 .
\end{lem}
\begin{proof}
Using Lemma  \ref{flow} (ii) and Lemma \ref{Iteration} (iv) and (v) this follows by induction on $n\in\N$.
	\extended{And here is the complete proof:
		For $n=1$ we have
 			\begin{align*}
 				P_{\calZ^\pm_1}(t) & \le  R_0+\int_0^t\Normli{F_0}(\sigma)\,d \sigma \quad\text{with Lemma \ref{flow}(iv)}\\
									&  \le R_0+\int_0^t C_1P_0^2(\sigma)+C_2 P_0^4(\sigma)\,d \sigma 
										\quad\text{with Lemma \ref{Iteration}(\ref{E_n}) and (\ref{D_n})}\\
									& \le R_0+C_3\int_0^t R_0^2+R_0^4\, d \sigma \\
									& \le R_0+C_3\int_0^t P_0^2(\sigma)+P_0^4(\sigma)\, d \sigma \\
									& \le P(t)
									\end{align*}		
	The same computation yields for the induction $n$ to $n+1$
		\begin{align*}
 				P_{\calZ^\pm_{n+1}}(t) & \le  R_0+\int_0^t\Normli{F_n}(\sigma)\,d \sigma \quad\text{with Lemma \ref{flow}(iv)}\\
									&  \le R_0+\int_0^t C_1P_n^2(\sigma)+C_2 P_n^4(\sigma)\,d \sigma 
										\quad\text{with Lemma \ref{Iteration}(\ref{E_n}) and (\ref{D_n})}\\
									& \le R_0+C_3\int_0^t P_n^2(\sigma)+P_n^4(\sigma)\, d \sigma \\
									& \le P(t)
									\end{align*}	
	}
\end{proof}

In the next steps we prove that a bound on the $p$-support  on an interval
$J\subset[0, \infty),$ $0\in J$, implies bounds on
all other quantities as well as the existence of a smooth solution of
(\ref{rVPRD}) on the interval $J$.
We assume that we are furnished with such an interval $J$ and a continuous
monotonously increasing positive 
function, called $P$ again,  such that for all $n\in\N$, $\varepsilon\in[0,\,1]$, $t\in J$ and $x,p\in\R^3$  
\begin{equation*}
	P_n(t)\le P(t)\,.
\end{equation*}
In fact, this is at least true for $J=\Ii{0}{a}$.
In addition we introduce the following helpful conventions. 
Henceforward, continuous
monotonously increasing positive functions functions,
 which can be calculated by the knowledge of $P$
and the basic constants $k, S_0$ and $R_0$ alone,  are
denoted by $C(\cdot)$, $C_1(\cdot), C_2(\cdot)$. Especially they are independent of 
$n\in\N$ and $\varepsilon\in[0,\,1]$. While $C_1(\cdot), C_2(\cdot)$ will be fixed functions, 
the definition of $C(\cdot)$ may change from
line to line, e.g. we have 
\[ C(t) = P(t) = P(t)\exp(C(t))+C(t). \] 

\begin{lem}\label{boundderivatives}
	For all $t\in J$ we have $\Normli{\px\varrho^\pm_n(t)}+\Normli{\px E_n(t)}\le C(t)$.
\end{lem}
\begin{proof} Using Lemma \ref{P-Lemma} and Lemma \ref{flow}(iii) we compute
	\begin{align*}
 		\norm{\px \varrho^\pm_n(t, x)} 
		&	\le\int_{\norm{p} \le P_n(t)} \norm{\px \left(f^{\circ,\pm}(\calZ^\pm_n(0;t, x, p)\right)}\,dp\\
		&	\le P^3(t) C \Normli{\px\calZ^\pm_n(0;t,\cdot,\cdot)}\\
		& 	\le C(t) \exp \left(\int_0^t 1+\Normli{\px E_{n-1}(s)}\,ds\right)
	\end{align*}
	Plugging this estimate together with Lemma \ref{Iteration}(\ref{rho_n}) in Proposition \ref{Batt}(\ref{dxE2})   we have
	\begin{align*}
 		\Normli{\px E_n(t)}	
		&	\le C(t) \left( 1+\ln^*\left(\exp	\left(\int_0^t\Normli{\px E_{n-1}(s)}\,ds\right)	\right)\right)\\
		& 	\le C_1(t) \left( 1+\int_0^t\Normli{\px E_{n-1}(s)}\,ds\right)\,,
	\end{align*}
	where we choose $C_1(\cdot)$ in such a way that additionally $C_1(0)\ge \Normli{\px E_0(0)}=\Normli{\px E_0(t)}$.
	We define $C_2(\cdot)$ as the solution of 
	$$ C_2(t)=C_1(t) \left(1+\int_0^tC_2(\sigma)\,d \sigma\right)\,.$$
	By induction on $n$ we conclude that
	\begin{align*}
 			\Normli{\px E_n(t)}	
			& 	\le C_2(t)\quad\text{and}\quad \\
			\Normli{\varrho^\pm_n(t)}
			&	\le C(t)
	\end{align*}
	for all $t\in J$ and all $n\in\N$.

\end{proof}
For $n\in\N$ we define
	$$\alpha_n(t) := \sup\setb{\norm{\calZ^\pm_{n+1} (s;t,x,p)-\calZ^\pm_{n} (s;t,x,p)}}{0\le s\le t,\, x,p\in\R^3}\,.$$

\begin{lem}\label{f_Cauchy}
	For all $ n\in\N$ and all $t\in J$ we have
	\begin{align}
		\Normli{f^\pm_{n+1}(t)-f^\pm_{n}(t)} &\le C\frac{C_3(t)^n}{n!}t^n\,\tag{i}\\
		\diss\alpha_n(t) &\le C(t)\frac{C_3(t)^{n-1}}{(n-1)!}t^{n}\,.\tag{ii}
	\end{align}
\end{lem}
\begin{proof}
	In the first step we compute
	\begin{equation}\label{f-Z-est}
		\norm{f^\pm_{n+1}(t, x, p)-f^\pm_{n}(t, x, p)} \le C\norm{\calZ^\pm_{n+1}(0;t,x,p)-\calZ^\pm_{n}(0;t,x,p)}\,.
	\end{equation}
	Using the characteristic equations we also have for all $0\le s\le t$ and all $x,p\in\R^3$
	\begin{align*}
 		\norm{\calX^\pm_{n+1}(s;t,x,p)-\calX^\pm_{n}(s;t,x,p)}
		&	\le \int_s^t \norm{\calP^\pm_{n+1}(\sigma;t,x,p)-\calP^\pm_{n}(\sigma;t,x,p)}\,d \sigma\,,\\
		\norm{\calP^\pm_{n+1}(s;t,x,p)-\calP^\pm_{n}(s;t,x,p)}
		& \le \int_s^t \norm{E_{n}(\sigma;\calX_{n+1}(\sigma))-E_{n-1}(\sigma;\calX_{n}(\sigma))}+
			\Big|\D _{n}(s)- \D _{n-1}(s)\Big|\,d \sigma\,.
	\end{align*}
	Utilizing Lemma \ref{boundderivatives} the electric fields can be estimated by
	\begin{align*}
 			\norm{E_{n}(\sigma;\calX_{n+1}(\sigma))-E_{n-1}(\sigma;\calX_{n}(\sigma))}
			& \le \Normli{\px E_{n}(\sigma)} \norm{\calX^\pm_{n+1}(\sigma)-\calX^\pm_{n}(\sigma)}
				+	\Normli{E_{n}(\sigma)-E_{n-1}(\sigma)}\\
			&	\le C(\sigma)\norm{\calX^\pm_{n+1}(\sigma)-\calX^\pm_{n}(\sigma)}
				+C(\sigma)\Normli{f^\pm_{n}(\sigma)-f^\pm_{n-1}(\sigma)}\,.
	\end{align*}
	Employing Lemma \ref{Calderon} we have for the dipols
	\begin{align*}
  		\norm{\D _{n}(\sigma) -\D _{n-1}(\sigma)}
	  	& 	\le  C\bigg(\Normlt{\varrho_{n}^+(\sigma) 
				-\varrho_{n-1}^+(\sigma)}\Normlt{j_{n}^-(\sigma)}
				+\Normlt{\varrho_{n-1}^+(\sigma)}\Normlt{j_{n}^-(\sigma)-j_{n-1}^-(\sigma)}\\
		&	\quad     
   			+ \Normlt{\varrho_{n}^-(\sigma) -\varrho_{n-1}^-(\sigma)}\Normlt{j_{n}^+(\sigma)}
			+\Normlt{\varrho_{n-1}^-(\sigma)}\Normlt{j_{n}^+(\sigma)-j_{n-1}^+(\sigma)} \bigg) \\  
		& \le C(\sigma)\Normli{f^\pm_{n+1}(\sigma)-f^\pm_{n}(\sigma)}.
	\end{align*}
Collecting these two estimates we conclude
	\begin{align*}
 		\norm{\calZ^\pm_{n+1}(s)-\calZ^\pm_{n}(s)}
		&	\le \int_s^tC(\sigma)\norm{\calZ^\pm_{n+1}(\sigma)-\calZ^\pm_{n}
		(\sigma)}+C(\sigma)\Normli{f^\pm_{n}(\sigma)-f^\pm_{n-1}(\sigma)}\,d \sigma
	\end{align*}
 		and using Gronwall's Lemma
		\begin{align}\label{Z-Gronwall}
			\norm{\calZ^\pm_{n+1}(s)-\calZ^\pm_{n}(s)}
			& 	\le \int_s^tC(\sigma)\Normli{f^\pm_{n}(\sigma)-f^\pm_{n-1}(\sigma)}\,d \sigma\,
				\exp \left(\int_s^tC(\sigma)\,ds\right)\,.
		\end{align}
	Combining (\ref{f-Z-est}) and (\ref{Z-Gronwall})
	 we have
	\begin{equation*}
		\Normli{f^\pm_{n+1}(t)-f^\pm_{n}(t)} \le C_3(t)\int_0^t\Normli{f^\pm_{n}(s)-f^\pm_{n-1}(s)}\,ds \,.
	\end{equation*}
	By induction it follows that
	\begin{align*}
 		\Normli{f^\pm_{n+1}(t)-f^\pm_{n}(t)} \le
		\sup\setb{\Normli{f^\pm_{1}(s)-f^\pm_{0}(t)}}{0\le s\le t}\frac{C_3(t)^n}{n!}t^n=C\frac{C_3(t)^n}{n!}t^n\,.	
	\end{align*}
	Plugging this inequality into  (\ref{Z-Gronwall}) we get
	\begin{equation*}
 		\alpha_n(t) \le C(t)\frac{C_3^{n-1}(t)}{(n-1)!}t^{n}.	
	\end{equation*}
	
	\end{proof}
	For a compact intervall $I\subset J$ we define
	$$\Delta_I:=\set{(s,t)}{t\in I, 0\le s\le t}\,.$$
	
\begin{cor}
	For every compact interval $I\subset J$ the sequences $(f^\pm_n)_n$, $(\calZ^\pm_n)_n$,  
	$(\varrho^\pm_n)_n$, $(j^\pm_n)_n$, $(E_n)_n$ and $(\D_n)_n$ are Cauchy sequences
	uniformly in $I\times\R^3\times\R^3$, resp. $\Delta_I\times\R^3\times^3$, resp. $I\times\R^3$, resp. $I$ and $\varepsilon\in[0,\,1]$.
\end{cor}
\begin{proof} This is an immediate consequence of Lemma \ref{f_Cauchy} in combination with Lemma \ref{Iteration}.
\end{proof}
\begin{lem}  The sequence $\left(\px E_n\right)_n$ is Cauchy sequence uniformly in $I\times\R^3$ and $\varepsilon\in[0,\,1]$
			for every compact interval $I\subset J$.
\end{lem}
	\begin{proof}
		Using Proposition \ref{Batt}(\ref{dxE2}) for every $0<d\le R$ we have
		\begin{align*}
			\Normli{\px E_n(t)-\px E_m(t)}
			& 	\le C \Big[R^{-3}\Normli{\varrho_n(t)-\varrho_m(t)}+d\Normli{\px\varrho_n(t)-\px\varrho_m(t)}\\
			&\quad	+\left(1+\ln \left(R/d\right)\Normlo{\varrho_n(t)-\varrho_m(t)}	\right)\Big] \,.
		\end{align*}
		Choosing $d$ sufficiently small proves the claim.
	\end{proof}
\subsection{$C^1$-solution and estimates}\label{C1solution}
	We define 
		\begin{align*}
			\lim_{n\to\infty}f^\pm_n 						&=:f^\pm\in C(J\times\R^3\times\R^3)
			&\lim_{n\to\infty} \calZ^\pm_n 				&=:\calZ^\pm\in C(\Delta_J\times\R^3\times\R^3;\R^3\times\R^3)\\
			\lim_{n\to\infty} \varrho^\pm_n 			&=:\varrho^\pm\in C(J\times\R^3)
			&\lim_{n\to\infty} j^\pm_n 					&=:j^\pm\in C(J\times\R^3;\R^3)\\
			\lim_{n\to\infty} E_n 							&=:E\in C^1(J\times\R^3;\R^3)
			&\lim_{n\to\infty}\D_n 						&=:\D\in C(J;\R^3)\,.
		\end{align*}
	
	\begin{theo}\label{C1solution-theo}
		The triple $(f^\pm, E, \D)$ is the unique $C^1$-solution of (\ref{rVPRD}) and for all $t\in J$ and $l\in \left\{0, 1\right\}$ we have
		\begin{align}
          		\tag{i} &\Normli{\ptxp^l f^\pm(t)}+\Normli{\ptx^l \varrho^\pm(t)}+
					\Normli{\ptx^l j^\pm(t)}+\Normli{\ptx^l E(t)}+\norm{\frac{d^l}{dt^l}\D(t)}\le C(t) \\
			\tag{ii} &f^\pm(t, x, p) = 0 \quad\text{if}\quad \norm{x}\ge X(t) 
					\quad\text{or}\quad \norm{p}\ge P(t)\,.		
		\end{align}
	\end{theo}
	\begin{proof}
		Clearly the densities $f^\pm(t)$, $ \varrho^\pm(t)$ and 
		$j^\pm(t)$ inherit the bounds on the supports $X(t)$ and $P(t)$. By standard theorems of integration we conclude
		$f^\pm=f^{\circ, \pm}(\calZ^\pm(0;\cdot, \cdot, \cdot))$, $\varrho^\pm=\varrho_{f^\pm}$, $j^\pm=j_{f^\pm}$,
		as well as $E=E_{\varrho^+}-E_{\varrho^-}$ and $\D= \D_{\varrho^+j^-}-\D_{\varrho^-j^+}$. We set
		$\widetilde{\calZ^\pm}:=\calZ_{\pm( E+	\varepsilon \D)}$
		By standard theorems of ODEs we have $\widetilde{\calZ^\pm}\in C^1(J\times J\times\R^3\times \R^3)$.
		Using the usual estimates for $0\le s \le t$, $t\in J$, $x, p\in\R^3$
		\begin{align*}
 			\norm{\calZ^\pm_n-\widetilde{\calZ^\pm}}(s;t,x,p)
			& \le \int_s^t 	\norm{\calP^\pm_n-\widetilde{\calP^\pm}}(\sigma)
								+\norm{E_n(\sigma, \calX^\pm_n(\sigma))-E(\sigma, \widetilde{\calX^\pm}(\sigma))}
								+\norm{\D_n-\D}(\sigma)\,d \sigma\\
			&\le \int_s^t (1+\Normli{\px E_n(\sigma)})\norm{\calZ^\pm_n-\widetilde{\calZ^\pm}}(\sigma)+
							\Normli{E_n(\sigma)-E(\sigma)}+\norm{\D_n-\D}(\sigma)\,d \sigma\\
			&\le \int_s^t C(\sigma)\norm{\calZ^\pm_n-\widetilde{\calZ^\pm}}(\sigma)\,d \sigma +\varepsilon_n(t)
		\end{align*}
		with a null sequence $\varepsilon_n(t)$.
		Using Gronwall's Lemma we conclude
		$$ \norm{\calZ^\pm_n-\widetilde{\calZ^\pm}}(s;t,x,p) \le \varepsilon_n(t)C(t)$$
		 and thus
		$$\calZ^\pm=\Restr{\widetilde{\calZ^\pm}}{\Delta_J\times\R^3\times\R^3}\in C^1(\Delta_J\times\R^3\times\R^3)\,.$$
		Therefore, $f^\pm$, $\varrho^\pm$, $j^\pm$ and  $\D$ are also $C^1$ on $J$. Using Lemma \ref{flow}(v)
		$(f^\pm, E, \D)$ is a solution of (\ref{rVPRD}). Now we shall establish bounds on the derivatives 
		independently of $\varepsilon\in [0,1]$.
		Employing Lemma \ref{Iteration} and Lemma \ref{boundderivatives} we already know that 
			$$\Normli{f^\pm(t)}+	\Normli{\varrho^\pm(t)}+\Normli{j^\pm(t)}
			+\norm{\D(t)}+\Normli{E(t)}+\Normli{\px E(t)}\le C(t)\,.$$
		Using Lemma \ref{flow}(iii) we also have that
		$$\Norm{\pxp\calZ^\pm(s;t, \cdot,\cdot)}\le C(t) $$
		for all $0\le s\le t$. But this implies $\Normli{\pxp f^{\pm}(t)}\le C(t)$. The bound on the $t$-derivative is established by means of
		the Vlasov equation, $\Normli{\pt f^\pm(t)}\le C(t)$. Utilizing the support properties we finally conclude
		$$\Normli{\ptx \varrho(t)}+\Normli{\ptx j^\pm(t)}+\Normli{\ptx E(t)}+\norm{\frac{d}{dt} \D(t)}\le C(t)\,.$$

	With regard to uniqueness we assume that we are established with a second solution $g^\pm$ according to 
	Definition \ref{solution-def} and $f^\pm(0)=g^\pm(0)$. 
	By Definition \ref{solution-def}(iii) and Lemma \ref{flow}  
	both solutions $f^\pm(t)$ and $g^\pm(t)$ are supported in a compact subset in $\R^6$ and 
	which can be chosen independently of $t\in I$ for every compact subinterval $I\subset J$.
	We replace the iterates $f^\pm_n$ and $f^\pm_{n-1}$ in Lemma \ref{f_Cauchy} by $f^\pm$ and $g^\pm $ and conclude
		$$\Normli{f^\pm(t)-g^\pm(t)}\le C_1(t)\int_0^t \Normli{f^\pm(s)-g^\pm(s)}\, ds$$
	and uniqueness follows.
	\end{proof}

	\begin{rem} The usual continuation property also holds for (\ref{rVPRD}): 
	an a-priori bound on the velocity support 
	yields global existence of solution. This can be shown in the same way as in \cite[Step 7, p. 401]{rein:07}.
	The crucial point is that the constants $C_1$ and $C_2$ in (\ref{P_def}) do only depend on the $L^1$ and the 
	$L^\infty$ norm  and not on the size of the support of the initial data.
	\end{rem}

\subsection{$C^k$-solutions and $C^k$-estimates}\label{Sec-k}	
	\begin{theo}\label{Theo-k}
		$f^\pm, \varrho^\pm, j^\pm, E$ and $\D$ are $C^k$ and for all integers $0\le j\le k$ we have the estimate
		\begin{equation}\label{Ck-estimate}
			\Normli{\ptxp^j f^\pm(t)}+\Normli{\ptx^j\varrho^\pm(t)}
			+\Normli{\ptx^l j^\pm(t)}+\Normli{\ptx^l E(t)}+\norm{\frac{d^j}{dt^j}\D(t)}\le C(t).
		\end{equation}
	\end{theo}
		We shall proof this theorem by induction on $1\le l \le k$.
		Before starting we  give a precise statement of the induction hypothesis.
		
		\begin{defi} For $1\le l \le k$  we say $\calH(l)$ holds true, iff the following holds:
		\begin{itemize}
			\item[(i)] 	For all $\quad 1\le j\le l$ we have 
						\begin{equation}\nonumber
							\Normli{\pxp^j \calZ^\pm_n(0;t,\cdot,\cdot)} 
							+\Normli{\pxp^j \varrho^\pm_n(t)}+\Normli{\px^j E_n(t)} \le C(t)
						\end{equation}
			\item[(ii)] 	For all $0 \le j\le l-1$ the sequences $(\pxp^j f^\pm_n)_n$, $(\pxp^j\calZ^\pm_n)_n$ 
							and $(\px^j \varrho^\pm_n)_n$ converge to $\pxp^j f^\pm$, $\pxp^j\calZ^\pm$ 
							and  $\px^j \varrho^\pm$ uniformly in $\varepsilon\in\II{0}{1}$ 
							and uniformly on  
							$I\times\R^3\times\R^3$, $\Delta_I\times\R^3\times\R^3$ and $I\times\R^3$, respectively,
							for all compact subintervals $I\subset J$.
			\item[(iii)] 	For all $0\le j \le l$, $(\px^j E_n)_n$ converges to
							$\px^j E$ uniformly in $\varepsilon\in\II{0}{1}$ and on $I\times\R^3$,  
							for all compact subintervals $I\subset J$.
			\item[(iv)]     The functions $f^\pm, \varrho^\pm, j^\pm, E$ and $\D$ are $C^l$ and for all integers $0\le j\le l$ 
							(\ref{Ck-estimate}) holds true.
	\end{itemize}
	\end{defi}
		\begin{proof}
		For $l=1$, $\calH(1)$ is already proved by means of  Theorem \ref{C1solution-theo}.
		Now we assume that $\calH(l)$ holds true for some $1\le l\le k-1$.
		We need a suitable representation of higher derivatives of composed functions. 
		Consider a multi-index $\alpha=(\alpha_x,\alpha_p)\in\N^3_0 \times \N^3_0$,  
		and a higher order partial derivative $\partial^\alpha = \partial_x^{\alpha_x}\,\partial_p^{\alpha_p}$. 
		Let $\map{G}{\R^3\times\R^3}{\R}$, resp. $\R^3$
		and $\map{H}{\R^3\times\R^3}{\R^3\times\R^3}$ be sufficiently smooth.
		By induction on $1 \le |\alpha|\le k$
		the following representation can be verified.
		\begin{equation}\label{Formel}
							\p^\alpha(G(H(x, p)) = \sum_{1\le|\beta|\le|\alpha| \atop \beta=(\beta_x, \beta_p)\in \N_0^3\times\N_0^3} 
								(\p^\beta G)(H(x, p))\cdot Q_{\alpha \beta}(\p^\gamma H)(x,p)
		\end{equation}
		where $ Q_{\alpha, \beta}$ are certain homogeneous polynomials of  degree
		$|\beta|$ in the variables $\partial^\gamma H$, where
		$\gamma=(\gamma_x, \gamma_p)$ varies in
		$\N_0^3 \times \N_0^3$ and
		$1\le|\gamma|\le|\alpha|+1-|\beta|$.
		In addition, if $|\beta|=1$, then $Q_{\alpha, \beta}$ is a polynomial
		in the variables $\partial^\gamma {H}$, $|\gamma|=|\alpha|$. 
		An explicit expression of these polynomials can be found in
		\cite{fraenkel:78}.

	 	In the first step we shall establish bounds on $\pxp^{l+1}\calZ^\pm_n$, $\px^{l+1}\varrho^\pm_n$ and $\px^{l+1} E_n$.
		By formula (\ref{Formel}) and $\calH(l)$ we estimate for $x\in\R^3$
			\begin{align}\nonumber
				\norm{\px^{l+1}\varrho^\pm_n(t, x)}
					&\le \int_{\norm{p}\le P(t)} \norm{\px^{l+1} \left(f^{\circ, \pm}(\calZ_n^{\pm}(0;t, x, p\right)}\, dp\\
					& \le C(t)+C(t)\Normli{\pxp^{l+1} \calZ^\pm_n(0;t,\cdot,\cdot)}\label{lp1rho-bound} 
			\end{align}
		Differentiating the characteristic equation 
		we compute for $0\le s\le  t$ and $x, p\in\R^3$, using the short forms  $(s)=(s; t , x,p)$ and
		$(\sigma)=(\sigma;t, x,p)$
		\begin{align}\label{Formel1}
 			\norm{\pxp^{l+1}\calZ^\pm_n(s)} 
			&\le \int_s^t \norm{\pxp^{l+1}\calP^\pm_{n}(\sigma)}+\norm{\pxp^{l+1}\left(E_{n-1}(\sigma, \calX^\pm_n(\sigma))\right)}\,d \sigma
		\end{align}
	With formula (\ref{Formel}) and the induction hypothesis $\calH(l)$ we estimate
		\begin{align}\nonumber
 			\norm{\pxp^{l+1}\left(E_{n-1}(\sigma, \calX^\pm_n(\sigma))\right)}\le C(\sigma)+	
		C(\sigma)\Normli{\px^{l+1}E_{n-1}(\sigma)}
			+	C(\sigma)\norm{\pxp^{l+1}\calZ^\pm_n(\sigma)}\,.
		\end{align}
	Furthermore, employing Lemma \ref{Batt}(i)  and Lemma \ref{P-Lemma} we also have
		\begin{align}\label{dlp1-E}
			\Normli{\px^{l+1}E_{n-1}(\sigma)} &\le C(\sigma)\Normli{\px^{l+1}\varrho^\pm_{n-1}(\sigma)}\,.
		\end{align}
	Inserting these two  estimates into (\ref{Formel1}) we have
		\begin{align}\nonumber
 			\norm{\pxp^{l+1}\calZ^\pm_n(s)} 
			&\le \int_s^tC(\sigma)\left(1+\Normli{\px^{l+1}\varrho^\pm_{n-1}(\sigma)}+\norm{\pxp^{l+1}\calZ^\pm_n(\sigma)}\right)\, d \sigma\,.
		\end{align}
	Thus, using Gronwall's Lemma,
		\begin{align}\label{Formel2}
 			\Normli{\pxp^{l+1}\calZ^\pm_n(0;t, \cdot,\cdot)} 
			&\le C(t)+C(t)\int_0^t \Normli{\px^{l+1}\varrho^\pm_{n-1}(\sigma)}\,d \sigma\,.
		\end{align}
	Plugging this inequality into (\ref{lp1rho-bound}) we arrive at
		\begin{align*}
 			\Normli{\pxp^{l+1}\varrho^\pm	_n(t)} 
			&\le C_4(t)+C_5(t)\int_0^t \Normli{\px^{l+1}\varrho^\pm_{n-1}(\sigma)} \, d \sigma\,,
		\end{align*}
		where we chose $C_4(\cdot)$ in such a way that $\Normli{\pxp^{l+1} f^{\pm,\circ}}\le C_4(0)$.
		We define $C_6(t)=C_4(t)\exp(tC_5(t))$. By induction on $n$\extended{\footnote{Details are explained in the appendix.\label{fn:1}}}
		 $$\Normli{\pxp^{l+1}\varrho^\pm	_n(t)} \le C_6(t)\quad\text{for all $n\in\N$}\,.$$
		Combining this with (\ref{dlp1-E}) and (\ref{Formel2}) proves $\calH(l+1)$(i).

	In the second step we proof $\calH(l+1)$(ii).
	Using formula (\ref{Formel}) we have
	\begin{align}\nonumber
 		\lefteqn{\norm{\pxp^l f^\pm_n(t, x, p)- \pxp^l f^\pm(t, x, p)}
		= \norm{\pxp^l \left(f^{\circ, \pm}(\calZ^\pm_n(0))\right)-\pxp^l \left(f^{\circ, \pm}(\calZ^\pm(0))\right)}}\\
		\nonumber
		&\le \sum_{\norm{\alpha}=l}\sum_{1\le\norm{\beta}\le l}
			\norm{\p^\beta f^{\circ, \pm}(\calZ^\pm_n(0))Q_{\alpha \beta}\left(\p^\gamma\calZ^\pm_n(0)\right)
			-\p^\beta f^{\circ, \pm}(\calZ^\pm(0))Q_{\alpha \beta}\left(\p^\gamma\calZ^\pm(0)\right)}\\
		\nonumber
		&\le \sum_{\norm{\alpha}=l}\sum_{1\le\norm{\beta}\le l}
			\norm{\p^\beta f^{\circ, \pm}(\calZ^\pm_n(0))-\p^\beta f^{\circ, \pm}(\calZ^\pm(0))}
			\norm{Q_{\alpha \beta}\left(\p^\gamma\calZ^\pm_n(0)\right)}\\
		\nonumber
		&\qquad\quad+\norm{\p^\beta f^{\circ, \pm}(\calZ^\pm(0))}\norm{Q_{\alpha \beta}\left(\p^\gamma\calZ^\pm_n(0)\right)-
				Q_{\alpha \beta}\left(\p^\gamma\calZ^\pm(0)\right)}\\
		&\le \varepsilon_n(t)C(t)+C\norm{\pxp^{l}\calZ^\pm_n(0)-\pxp^l\calZ^\pm(0)}\label{fnl-Cauchy}
	\end{align}
	with a null sequence $\varepsilon_n(t)$ which is independent of $\varepsilon\in\II{0}{1}$.
	Next we have to estimate the derivatives of the characteristics:
	Using the abbreviation 
	$\zeta_{n,l}(s):=\norm{\pxp^l\calZ^\pm_n(s;t,x,p)-\pxp^l\calZ^\pm(s;t,x,p)}$
	 we compute 
	\begin{align*}
 			\zeta_{n,l}(s)
			&\le\int_s^t \zeta_{n,l}(\sigma)+
			\normlr{\pxp^l \left(E_{n-1}(\sigma, \calX^\pm_n(\sigma))\right)
			-\pxp^l\left(E(\sigma, \calX^\pm(\sigma))\right)}\,d \sigma\,.
	\end{align*}
	With formula (\ref{Formel}) we have for the second sumand in the integral
	\begin{align*}
 		\lefteqn{\normlr{\pxp^l \left(E_{n-1}(\sigma, \calX^\pm_n(\sigma))\right)
			-\pxp^l\left(E(\sigma, \calX^\pm(\sigma))\right)}}\\
		& \le \sum_{\norm{\alpha}=l}\sum_{1\le\norm{\beta}\le l}
			\normlr{\p^\beta E_{n-1}(\calZ^\pm_n)Q_{\alpha, \beta}(\p^\gamma \calZ^\pm_n)-
						\p^\beta E(\calZ^\pm)Q_{\alpha, \beta}(\p^\gamma \calZ^\pm)}\\
		&\le \sum_{\norm{\alpha}=l}\sum_{1\le\norm{\beta}\le l}
				\normlr{\p^\beta E_{n-1}(\calZ^\pm_n)-\p^\beta E_{n-1}(\calZ^\pm)}
				\norm{Q_{\alpha, \beta}(\p^\gamma \calZ^\pm_n)}\\
		&\qquad\qquad\qquad+ \left. \normlr{\p^\beta E_{n-1}(\calZ^\pm)-\p^\beta E(\calZ^\pm)}
			\normlr{Q_{\alpha, \beta}(\p^\gamma \calZ^\pm_n)}\right.\\
		&\qquad\qquad\qquad+  \normlr{\p^\beta E(\calZ^\pm)}
		\normlr{Q_{\alpha, \beta}(\p^\gamma \calZ^\pm_n)-Q_{\alpha, \beta}(\p^\gamma \calZ^\pm)}
		\end{align*}
 	With regard to the summands in the first line we have
	$$\normlr{\p^\beta E_{n-1}(\calZ^\pm_n)-\p^\beta E_{n-1}(\calZ^\pm)}
				\norm{Q_{\alpha, \beta}(\p^\gamma \calZ^\pm_n)}
		\le \Normli{\px^{\norm{\beta}+1}E_{n-1}(\sigma)}\normlr{\calZ^\pm_n(\sigma)-\calZ^\pm(\sigma)}
			\norm{Q_{\alpha, \beta}(\p^\gamma \calZ^\pm_n)}\le \varepsilon_n(t)C(t)\,.$$
		Observe that we already have established the bound on $\Normli{\px^{l+1}E_n}$.
		For the second line we have
	$$\normlr{\p^\beta E_{n-1}(\calZ^\pm)-\p^\beta E(\calZ^\pm)}
			\normlr{Q_{\alpha, \beta}(\p^\gamma \calZ^\pm_n)}
		\le\Normli{\px^{\norm{\beta}}E_{n-1}(\sigma)-\px^{\norm{\beta}}E(\sigma)}
		\normlr{Q_{\alpha, \beta}(\p^\gamma \calZ^\pm_n)} \le \varepsilon_n(t)C(t)\,.$$
		In the third line summands with $2\le\norm{\beta}\le l$ can be estimated against $\varepsilon_n(t)C(t)$,
		whereas summands with $\norm{\beta}=1$ can be estimated against $C(\sigma)\zeta_{n,l}(\sigma)$.
		Collecting everything we conclude
	\begin{equation*}
 		\zeta_{n, l}(s)\le \varepsilon_n(t)	C(t)+	C(t)\int_s^t \zeta_{n,l}(\sigma)\,d \sigma
	\end{equation*}
		and using Gronwall's Lemma
	\begin{equation*}
 			\zeta_{n, l}(0)\le \varepsilon_n(t)	C(t).
	\end{equation*}
	Combining this inequality with (\ref{fnl-Cauchy}) proves $\calH(l+1)$(ii).

	For the third step we apply Lemma \ref{Batt}(iii) on $\px^{l}E_n$ and $\px^l \varrho_n$ and chose  $d$ sufficiently small again.
	Thus, $\calH(l+1)$(iii) is shown.

	For the last step we note  that by formally differentiating the characteristic equations
	\begin{eqnarray*}
 		\ptxp^{l+1} \calX^{\pm}(s) & = & \ptxp^{l+1} \left(x+\int_t^s\calP^\pm(\sigma)\, d \sigma\right)\\
		\ptxp^{l+1}\calX^{\pm}(s)& = & \ptxp ^{l+1}\left(p+\int_t^s E(\sigma, \calX^\pm(\sigma))+\D(\sigma)\, d \sigma\right)
	\end{eqnarray*} 
	we end with an initial value problem for an  inhomogeneous linear ODE in $\ptxp^{l+1}\calZ^\pm$
	$$\frac{d}{ds}\ptxp^{l+1}\calZ^\pm = A(s, t,x, p)\ptxp^{l+1}\calZ^\pm+B(s;t,x,p)\,\qquad \ptxp^{l+1}\calZ^\pm(t;t,x,p)=D(t,x,p)$$
	where the coefficient functions are continuous and 
	$$ \norm{A(s, t,x, p)}+\norm{B(s, t,x, p)}+	\norm{D(t,x,p)}\le C(t)$$
	for all $0\le s\le t $ and $x,p\in\R^3$. For this step it is crucial that the only derivates of 
	order $l+1$ applied on $E$ and $\D$ are purely spatial and thus continuous.
	Therefore we have $\calZ^\pm\in C^{l+1}(J\times J\times\R^3\times\R^3)$ and 
	using a standard estimate of ODEs, see e.g. \cite[16.VI]{walter:98},
	$$\norm{\ptxp^{l+1}\calZ^{\pm}(s)}\le C(t). $$
	Using this together with the support properties of $f^\pm$ gives $\calH(l+1)$(iv) and Theorem
	\ref{Theo-k} is completely proved.
	\end{proof}
\extended{
\begin{appendix}
\section{The Gronwall type estimate in footnote \ref{fn:1}}
Fix some $\bar{t}\in J$. For all $0\le t\le\bar{t}$ we have
	$$\Normli{\px^{l+1}\varrho_n^\pm(t)} \le C_4(\bar{t})+	
			C_5(\bar{t})\int_0^t\Normli{\px^{l+1}\varrho_{n-1}^\pm(\sigma)}\, d \sigma\,.$$
	Define $y_{\bar{t}}(t)=C_4(\bar{t})\exp(tC_5(\bar{t})$. This is the solution of the equation
	$$ y_{\bar{t}}(t)=C_4(\bar{t})+	
			C_5(\bar{t})\int_0^ty_{\bar{t}}(\sigma)\, d \sigma\,.$$
	By induction on $n$ it follows that for all $n\in\N$ and all $0\le t\le \bar{t}$
	$$\Normli{\px^{l+1}\varrho_n^\pm(t)} \le y_{\bar{t}}(t)\le C_6(\bar{t})\,,$$
	and especially $\Normli{\px^{l+1}\varrho_n^\pm(\bar{t})}\le C_6(\bar{t})$.
\section{Zur Energiegleichung (\ref{StEne-def})}
\subsection{Skalierungsargumente}
Wir zeigen durch ein Skalierungsargument, dass keine bessere Ungleichung m\"oglich ist, die $M_1(f)$ gegen 
$M_2(f)$ und weitere Normen von $f$ absch\"atzt. Hierbei definieren wir $M_k(f) = \iint \norm{p}^k f(x, p)\,dp\,dx$.
Dazu definieren wir f\"ur $\lambda, \mu\in\R_+$ die skalierte Funktion
$f_{\lambda, \mu}(x, p)=f(\lambda x, \mu p)$ und berechnen:
\begin{eqnarray*}
 M_1(f_{\lambda, \mu}) & = & \iint p f_{\lambda, \mu}(x, p)\,dp\,dx = \iint p f(\lambda x, \mu p)\,dp\,dx \\
				& = & \lambda^{-3}\mu^{-4} \iint wf(y, w)\,dw\,dy\\
				& = & \lambda^{-3} \mu^{-4} M_1 (f) \\
\Normlp{f_{\lambda, \mu}} & = & \left(\iint \norm{f(\lambda x, \mu p)}^p\,dp\,dx\right)^{1/p}\\
				& = & \left(\lambda^{-3}\mu^{-3}\iint\norm{f(y, w)}^p\,dw\,dy\right)^{1/p}\\
				& = & \lambda^{-3/p}\lambda^{-3/p}\Normlp{f}\\
\Normlp{\varrho_{f_{\lambda,\mu}}} & = & \left(\int\norm{f(\lambda x, \mu p)\,dp}^p\,dx\right)^{1/p}\\
			& = & \left(\int\norm{\mu^{-3}\int f(\lambda x, w)\,dw}^p\,dx\right)^{1/p}\\
			& = & \mu^{-3} \left(\int\norm{\int f(\lambda x, w)\,dw}^p\right)^{1/p}\\
			& = & \mu^{-3}\left(\lambda^{-3}\int\norm{\int f(y, w)\,dw}^p\right)^{1/p}\\
			& = & \mu^{-3} \lambda^{-3/p} \Normlp{\varrho_f}
\end{eqnarray*}
Durch Vergleich der Exponenten von $\lambda$ und $\mu$ sehen wir dass der Ansatz
$$ M_1(f)\le C \Normlp{f}^q M_2(f)^{1-q}$$
sofort auf $p=1$ und $q=1/2$ f\"uhrt, die Ungleichung die bereits etabliert ist.
Genauso erhalten wir mit dem Ansatz
$$ M_1(f)\le C \Normlp{\varrho_f}^q M_2(f)^{1-q}$$
wieder $p=1$ und $q=1/2$. Das ist nat\"urlich auch nicht \"uberraschend.
Wir k\"onnen uns vordergr\"undig mehr Freiheit durch den folgenden Ansatz verschaffen
$$  M_1(f)\le C \Normlp{\varrho_f}^{r} \Norm{\varrho_f}_q^s M_2(f)^{1-r-s}.$$
Anschlie\ss{}end m\"usste $\Norm{\varrho_f}_q$ aber auch wieder gegen $\Norm{f}_l^sM_2(f)^{1-s}$ abgesch\"atzt 
werden was uns insgesamt auf den Ansatz
$$ M_1(f)\le C \Normlp{f}^{r} \Norm{f}_q^s M_2(f)^{1-r-s}$$
f\"uhrt. Durch Vergleich der Exponenten sehen wir aber wieder, dass $p=q=1$ gelten muss, denn die $\lambda$-Gleichung
liefert gerade
$$ 3 = \frac{3r}{p}+\frac{3s}{q}+3(1-r-s)\,.$$
Durch das Skalierungsargument erkennt man auch sehr leicht, dass die höchste $L^p$-Norm von $\varrho_f$ die sich gegen eine
$L^{q}$-Norm von $f$ und $M_2(f)$ absch\"atzen l\"asst gerade die $5/3$-Norm ist:
Der Ansatz
$$ \Normlp{\varrho_f} \le C \Norm{f}_q^r M_2(f)^{1-r}$$ f\"uhrt
auf die beiden Gleichungen (erst f\"ur $\lambda$ dann f\"ur $\mu$)
\begin{eqnarray*}
 	\frac{3}{p} & = & \frac{3r}{q}+3(1-r)\\
	3			& = & \frac{3r}{q}+5(1-r)\,.
\end{eqnarray*}
Aus der $\mu$-Gleichung folgt $q= \frac{3r}{5r-2}$ und daraus $r\ge 2/5$ und f\"ur $r=2/5$ haben wir gerade $q=\infty$ und
mit der $\lambda$-Gleichung $p=5/3$.
\end{appendix}}
\bibliographystyle{plain} 
\bibliography{/Users/sebastianbauer/Dropbox/Uni/paper/paule,/Users/sebastianbauer/Dropbox/Uni/paper/Litbank}
\end{document}